\documentclass[conference]{IEEEtran}
\pdfoutput=1

\usepackage{amsmath}
\usepackage{algorithmic}
\usepackage{graphicx}
\usepackage{svg}
\usepackage[skip=3pt]{caption}
\usepackage{array}
\usepackage{cite}
\usepackage{hyperref}
\usepackage{xcolor,soul}
\usepackage{amssymb}

\usepackage{enumitem}
\usepackage{booktabs}
\usepackage{soul, color}
\usepackage{balance}

\usepackage{tikz}
\newcommand*\circled[1]{\tikz[baseline=(char.base)]{\node[shape=circle,draw,inner sep=0.5pt] (char) {#1};}}

\usepackage{booktabs, makecell, multirow, tabularx} 
\newcolumntype{C}{>{\centering\arraybackslash}X}
\newcolumntype{L}{>{\hsize=.4\hsize}C}
\newcolumntype{M}{>{\hsize=.35\hsize}C}
\newcolumntype{S}{>{\hsize=.25\hsize}C}

\usepackage{colortbl}
\definecolor{light-gray}{gray}{0.85}
\usepackage{caption} 
\usepackage{diagbox}
\usepackage{threeparttable}

\newenvironment{para_noindent}{\setlength{\parindent}{0pt}}{}
\definecolor{Gray}{gray}{0.9}

\definecolor{moh_colour}{RGB}{255, 204, 204}

\definecolor{yuz_colour}{RGB}{191, 232, 255}

\newcommand{\tb}[1]{\textbf{#1}}

\begin{document}

\title{
    Kratos: An FPGA Benchmark for Unrolled DNNs with Fine-Grained Sparsity and Mixed Precision
}


\author{
    Xilai Dai, Yuzong Chen, Mohamed S. Abdelfattah \\
    \textit{Department of Electrical and Computer Engineering, Cornell University}  \\
    \{xd44, yc2367, mohamed\}@cornell.edu
}
    

\maketitle

\begin{abstract}
FPGAs offer a flexible platform for accelerating deep neural network (DNN) inference, particularly for non-uniform workloads featuring fine-grained unstructured sparsity and mixed arithmetic precision. 
To leverage these redundancies, an emerging approach involves partially or fully \textit{unrolling} computations for each DNN layer.
That way, parameter-level and bit-level ineffectual operations can be completely skipped, thus saving the associated area and power.
Regardless, unrolled implementations scale poorly and limit the size of a DNN that can be unrolled on an FPGA.
This motivates the investigation of new reconfigurable architectures to improve the efficiency of unrolled DNNs, while taking advantage of sparsity and mixed precision.
To enable this, we present \textbf{Kratos}: a focused FPGA benchmark of unrolled DNN primitives with varying levels of sparsity and different arithmetic precisions.
Our analysis reveals that unrolled DNNs can operate at very high frequencies, reaching the maximum frequency limit of an Arria 10 device. 
Additionally, we found that substantial area reductions can be achieved through fine-grained sparsity and low bit-width.
We build on those results to tailor the FPGA fabric for unrolled DNNs through an architectural case study demonstrating $\sim$2$\times$ area reduction when using smaller LUT sizes within current FPGAs.
This paves the way for further exploration of new programmable architectures that are purpose-built for sparse and low-precision unrolled DNNs.
Our source code and benchmark are available on \href{https://github.com/abdelfattah-lab/Kratos-benchmark}{github.com/abdelfattah-lab/Kratos-benchmark}.
\end{abstract}

\section{Introduction}

Deep Neural Network (DNN) inference has become one of the most important compute workloads of our time, spanning many applications from image \cite{alexnet, resnet, mobileNet, mobileNetV2, vit} and speech recognition \cite{rnn} to natural language processing \cite{attention, bert, GPT2} and autonomous driving \cite{yolo, yolov3, yolox, ssd}. 
GPUs and custom ASIC chips currently dominate DNN inference, particularly because of their high compute capacity and memory bandwidth.
This enables very efficient dense matrix multiplication on these platforms.
However, it has been shown, time and again, that DNNs exhibit very high levels of \textit{fine-grained} sparsity~\cite{han:nips15:pruning} and can tolerate low and mixed arithmetic precision~\cite{kuan:cvpr19:haq}---two intrinsic properties that are challenging to accelerate on existing DNN accelerators.
This begs the question of whether there are more suitable architectures for sparse and low-precision DNNs.

FPGAs provide an attractive acceleration platform because of their high flexibility and bit-level programmability.
However, the reconfigurability overhead is generally very high, making FPGAs approximately an order of magnitude less efficient when compared to an ASIC implementation~\cite{kuon:tcad07:cad, boutros:trets18:cnn_gap}.
Even though many innovative and sparse-aware DNN accelerator architectures were introduced on FPGAs~\cite{han:fpga17:ese,abdelfattah:fpl18:dla,fan:micro22:fabnet,Meng2021FixyFPGAEF,Lu2019AnEH,Cao2019EfficientAE}, none gained enough traction to compete with current GPUs or ASICs. 
Nevertheless, there is an emerging \textit{style} of DNN acceleration on FPGAs that holds promise.
Specifically, \textbf{unrolled} DNN implementations, wherein a DNN accelerator contains partially or fully unrolled computation engines that are specialized for each DNN layer.

\begin{figure*}
    \centering
    \includegraphics[width=1\linewidth]{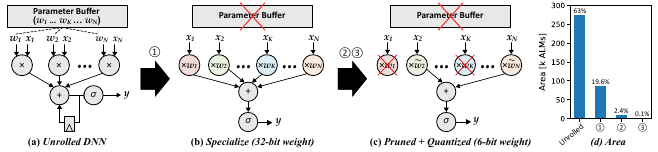}
    \caption{Diagram of unrolled DNNs and the area of a 64×64 matrix multiplication on an FPGA. Naïve unrolling quickly utilizes most of the FPGA area (63\%), but specialization \protect\circled{\small1}, pruning \protect\circled{\small2}, and quantization \protect\circled{\small3} reduce area by 600$\times$ down to just 0.1\% of the FPGA for 4096 effective FLOPs.}
    \vspace{-5pt}
    \label{fig:unrolled_dnn}
\end{figure*}

Fig.~\ref{fig:unrolled_dnn} shows a conceptual diagram of unrolled DNNs and the area of a 64$\times$64 matrix multiplication on an Arria~10 GX 1150 FPGA~\cite{arria10}. 
Full unrolling means having a hardware multiply-accumulate (MAC) unit for each MAC operation in the matrix multiplication, as shown in Fig. \ref{fig:unrolled_dnn}(a).
This na\"ive unrolling quickly utilizes most of the FPGA area (63\%) as shown in Fig. \ref{fig:unrolled_dnn}(d).
However, we consider unrolled DNN implementations that are specialized, pruned, and quantized.
Specialization of MAC units means converting them to multiply with constant weight parameters, as shown in Fig. \ref{fig:unrolled_dnn}(b). 
This drastically reduces compute area ($\sim$4$\times$) by optimizing the MAC circuitry and by leveraging bit-level sparsity within the parameter values.
Combining specialization with pruning and quantization, as shown in Fig. \ref{fig:unrolled_dnn}(c), further reduces area by $\sim$150$\times$ down to just 0.1\% of the FPGA for 4096 \textit{effective} FLOPs, making unrolled DNNs practical on current FPGAs.
Indeed, there are a number of recent works that successfully leverage this implementation methodology of unrolled DNNs on FPGAs, especially for smaller DNNs with very high throughput requirements~\cite{unuroglu:fccm20:logicnet,wang:tcomp20:lutnet,wang:fpga22:shrinkage,lou:trets23:fsead,nazemi:fccm21:nullanet,tridgell:trets19:ternary,duarte:jinst18:hls4ml}.

%

A key advantage of unrolled DNNs on FPGAs is the proportional reduction in circuit area and efficiency gains from all forms of redundancy. Conventional DNN accelerators, including GPUs, achieve only $\sim$15\% performance/watt improvement from 50\% \textit{structured} sparsity, even with dedicated hardware support~\cite{nvidia_sparse}, far short of the expected 2$\times$. FPGA's bit-level reconfigurability accelerates fine-grained and unstructured sparsity effectively. However, capacity limits exist: a 4-MFLOP DNN can be fully unrolled on an Arria~10 GX 1150 FPGA, as shown in Fig.~\ref{fig:unrolled_dnn}, but this is suitable only for small DNNs, highlighting the need for FPGA architectural exploration to enhance efficiency.

Open-source CAD and architecture exploration frameworks like VTR~\cite{vtr8} enable us to examine architectural tradeoffs, including LUT sizes, interconnection flexibility, and new hard blocks. We can prototype new programmable devices based on FPGAs, specifically designed to accelerate unrolled DNNs. This motivates our work on \textbf{Kratos}\footnote{Kratos personifies strength in Greek mythology. Our benchmark leverages the \textbf{strength} of FPGAs, specifically bit-level programmability, to accelerate sparse DNNs.}: a benchmark suite for unrolled DNNs with unstructured sparsity and mixed precision.

More specifically, we make the following contributions:

\begin{enumerate}
    \item We introduce the Kratos benchmark. A circuit benchmark suite of unrolled convolutional and general matrix multiplication (GEMM) DNN layers with different levels of fine-grained sparsity and numerical precisions.
    \item Unlike other FPGA benchmarks, our SystemVerilog code is human-readable, parameterized, and extensible, in addition to being compatible with both commercial (Quartus Prime) and academic (VTR) CAD flows.
    \item We present area and delay characteristics of our benchmark, showing that fully-unrolled DNNs can far exceed the clock network limitations on Arria 10 FPGAs. We also observe \textit{linear} improvements in efficiency with respect to higher sparsity and lower bitwidth.
    \item We perform an FPGA architectural case study investigating the most efficient LUT size for unrolled DNNs. We show $\sim$2$\times$ FPGA area reduction by tailoring FPGA logic blocks for unrolled DNNs, paving the way for the investigation of new purpose-built devices for sparse and low-precision DNN acceleration.
\end{enumerate}

\vspace{5pt}

While existing FPGA benchmarks provide valuable insights for general-purpose applications, they fall short in addressing the specific needs of unrolled DNNs with fine-grained sparsity and mixed precision. Kratos fills this gap by offering a specialized benchmark suite and related tools that enable architectural exploration and optimization for unrolled DNNs. This is crucial for designing next-generation programable accelerators that can fully leverage the potential of unrolled DNNs, achieving higher throughput and efficiency compared to traditional dense execution models.

\section{Related Work}

\textbf{Unrolled DNNs} follow the synchronous dataflow design paradigm, where DNN layers are partially or fully unrolled on an FPGA to match throughput between layers~\cite{venieris:ffcm16:fpgaconvnet,finn}. This approach efficiently implements binary/ternary DNNs~\cite{tridgell:trets19:ternary}, high-throughput data analysis~\cite{duarte:jinst18:hls4ml}, and anomaly detection~\cite{lou:trets23:fsead}. Recent work focuses on tailoring fully unrolled DNNs to FPGAs using LUT primitives, low arithmetic precision, and high unstructured sparsity~\cite{wang:tcomp20:lutnet,unuroglu:fccm20:logicnet,wang:fpga22:shrinkage}, achieving high efficiency compared to traditional DNN accelerators~\cite{abdelfattah:fpl18:dla}.

\textbf{FPGA Benchmark} circuits have commonly been used to guide the architecture exploration of FPGAs~\cite{vtrbench}.
Traditionally, a variety of benchmarks from different domains are used~\cite{Murray2013TitanEL} to maintain the general-purpose nature of FPGAs. Recently, some DNN-focused benchmarks have addressed the need for domain-specific FPGA fabrics for DNN acceleration. Koios~\cite{arora:fpl21:koios} includes DNN accelerator circuits with varied implementations, and Roorda et al.~\cite{roorda:trets22:dnnbench} released a flexible, auto-generated DNN benchmark suite for new DSP architecture investigation. \textbf{Kratos} focuses on (1) unrolled DNN implementations, (2) unstructured sparsity and mixed precision, and (3) enhancing FPGA logic and routing architecture.

\textbf{DNN-Optimized FPGAs} have been proposed by improving logic blocks for low-precision DNNs~\cite{luxor,boutros:fpga19:math-hard}, enhancing DSPs with more low-precision computation~\cite{boutros:fpl18:embrace-diversity,pir-dsp}, or augmenting BRAMs with compute capabilities~\cite{bramac,comefa,ccb,m4bram}.
We aim to use \textbf{Kratos} for investigating optimized logic block and routing architectures to create new domain-specific programmable devices for enhanced unrolled DNN performance. 

%
%

\section{Benchmark Description}

\subsection{Kernels}
The Kratos benchmark contains $8$ kernels as shown in Table \ref{tab:benchmark_summerization}. These kernels implement two main DNN operations: GEMM and convolution, which are heavily used by a wide range of DNNs. The GEMM operation is used by the fully-connected layer which is ubiquitous in many DNNs such as long short-term memory \cite{rnn} and transformers \cite{attention}, while the convolution operation dominates convolutional neural networks \cite{alexnet}. Since Kratos focuses on unrolled DNNs, weights are embedded into circuit connections like LUTs rather than memory. For instance, during multiplication, the input goes directly into LUTs, producing the output without needing to access a multiplier or load weights from BRAM.

\begin{table}
    \footnotesize	
    \centering
    \caption{The Kratos Benchmarks.}
    \renewcommand{\arraystretch}{1.1}

    \begin{threeparttable}
    \begin{tabular}{ m{1cm}  m{1.8cm} m{1.8cm} m{1.3cm} m{1.8cm}  }
        \toprule
          \textbf{Kernel} & \textbf{Unrolling\newline Factor} & \textbf{Input / \newline Cycle} & \textbf{Weight \newline Duplicate} & \textbf{Output / \newline Cycle} \\
        \bottomrule
          gemmt & row-parallel & $1 \times n$ & -- & $1 \times p$ \\
          \rowcolor{Gray} gemmt & fully-unrolled & $m \times n$ & $m\times$ & $n \times p$ \\
          gemms & row-parallel & $1 \times n$ & -- & $1 \times p$\\ 
          \rowcolor{Gray} conv1d & pixelwise & $F_w \times 1 \times I_c$ & -- & $1 \times 1 \times O_c$ \\
          conv1d & fully-unrolled & $I_W \times 1 \times I_c$ & $O_w\times$ & $O_w \times 1 \times O_c$ \\
          \rowcolor{Gray} conv2d & pixelwise & $F_w \times F_h \times I_c$ & -- & $1 \times 1 \times O_c$ \\
          conv2d & row-parallel & $I_W \times F_h \times I_c$ & $O_w\times$ & $O_w \times 1 \times O_c$ \\
          \rowcolor{Gray} conv2d & fully-unrolled & $I_W \times I_h \times I_c$ & $O_wO_h\times$ & $O_w \times O_h \times O_c$ \\
        \bottomrule
    \end{tabular}
    \begin{tablenotes}
      \item[1] All kernels accept user-defined sparsity $\in [0, 1]$ and precision $\in \mathbb{N}_{>0}$.
      \item[2] gemmt = multiply-add tree implementation of GEMM.
      \item[3] gemms = weight-stationary systolic implementation of GEMM.
    \end{tablenotes}
    \end{threeparttable}
    \vspace{-5pt}
    \label{tab:benchmark_summerization}
\end{table}

\begin{figure}
    \centering
    \includegraphics[width=1\linewidth]{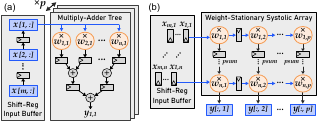}
    \caption{Hardware implementation of GEMM: (a) multiply-adder tree and (b) weight-stationary systolic array.}
    \label{fig:unroll_hardware}
\end{figure}

The GEMM dataflow is shown in Fig. \ref{fig:unroll_factor}(a), where the input matrix $x^{m \times n}$ is multiplied by the unrolled weight matrix $w^{n \times p}$ to generate the output matrix $y^{m \times p}$. Kratos contains two types of hardware implementations for GEMM that use multiply-adder tree (\textit{gemmt}) and weight-stationary systolic array (\textit{gemms}) as shown in Fig. \ref{fig:unroll_hardware}(a) and (b), respectively. The datapath of our design is heavily pipelined by inserting registers between every stage of multiplication or addition. For convolution, Kratos contains 1-D convolution (\textit{conv1d}) and 2-D convolution (\textit{conv2d}) implemented using the multiply-adder tree. Fig. \ref{fig:unroll_factor}(b) shows the dataflow of conv2d, where the $I_W \times I_h \times I_c$ input feature map is convolved with the $F_w \times F_h \times I_c \times O_c$ filter matrix to generate the $O_w \times O_h \times O_c$ output feature map. The conv1d kernel has a similar dataflow as conv2d except that $I_h = O_h = F_h = 1$. Using multiply-adder trees allows pruning leaves of zero weights while the traditional systolic array still needs structural registers to keep systolic and thus leads to low resource efficiency.

\subsection{Input Unrolling Factors}
\begin{figure}
    \centering
    \includegraphics[width=1\linewidth]{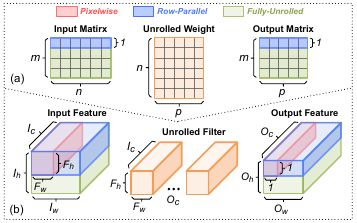}
    \caption{Dataflow of (a) GEMM and (b) convolution for different input unrolling factors: pixelwise, row-parallel, and fully-unrolled. The weight/filter is always fully unrolled}
    \label{fig:unroll_factor}
\end{figure}

All weights are fully unrolled in the Kratos kernels to take full advantage of parameter-level sparsity.
An important design consideration is the input unrolling factor---this quantifies the portion of the input tensor that can be processed simultaneously, and directly affects the resulting throughput. 
Kratos supports three input unrolling factors as illustrated in Fig. \ref{fig:unroll_factor} and described below. The different color boxes indicate the number of elements processed in one cycle. This visual representation helps to understand the efficiency gains achieved through our approach.

\begin{para_noindent}
  \textbf{Pixelwise}: This unrolling factor is applicable to convolution. As shown in Fig. \ref{fig:unroll_factor}(b), the pixelwise unrolling generates one pixel along all channels of the output feature map in parallel. 
  
  \textbf{Row-Parallel}: This unrolling factor is applicable to both GEMM and convolution. The row-parallel unrolling generates one row of the output matrix for GEMM, and one row along all channels of the output feature map for convolution in parallel as shown in Fig. \ref{fig:unroll_factor}(a) and (b), respectively. 
  
  \textbf{Fully-Unrolled}: For fully-unrolled GEMM, the whole output matrix can be generated in one shot as shown in Fig. \ref{fig:unroll_factor}(a). The fully-unrolled 1-D convolution is the same as the row-parallel 1-D convolution since $O_h = 1$. For fully-unrolled 2-D convolution, the entire input feature is processed simultaneously to obtain the whole output feature map in one shot as shown in Fig. \ref{fig:unroll_factor}(b). 
\end{para_noindent}

The unrolling factor impacts hardware design and resource utilization. For instance, the row-parallel \textit{gemmt} implementation broadcasts one row of the input matrix to the unrolled weight, generating one row of the output matrix per cycle. Input and weight duplication can improve throughput by processing more inputs in parallel. In fully unrolled implementations, the entire input matrix is processed simultaneously, obtaining the whole output matrix in $1$ cycle, with a throughput of $m \times n \times p$ operations. However, for \textit{gemms}, this results in diminishing returns due to the systolic propagation penalty~\cite{Eckert2023EideticAI}. For convolution, weight duplication is necessary for row-parallel and fully-unrolled implementations. We use BRAM for pixelwise unrolling and a shift-register network for row-parallel and fully-unrolled kernels to ensure sufficient input bandwidth.

\subsection{CAD for FPGA Architecture Exploration}
One of the main motivations of this work is to evaluate existing FPGA architectures and explore new optimized architectures for unrolled DNNs. 
To achieve this, Kratos is designed to be compatible with both the commercial Intel Quartus Prime and the open-source VTR~\cite{vtr8} flow. 

Creating a VTR-compatible benchmark has long been a labor-intensive process due to the limited Verilog syntax coverage of VTR's Odin II synthesis front-end \cite{OdinII}. However, Odin II provides efficient partial technology mapping for balancing soft logic and hard blocks of a target FPGA architecture. Recently, VTR has integrated Yosys \cite{Yosys}, an open-source synthesis tool with extensive Verilog-2005 and SystemVerilog support such as the "generate" statement. The new VTR synthesis front-end using a combination of Yosys for synthesis and Odin II for partial mapping \cite{Damghani2022YosysOdinIITO} significantly reduces the efforts of handling unsupported Verilog syntax. Hence, the Kratos benchmark uses this newly released VTR flow. 

\begin{figure}
    \centering
    \includegraphics[width=1\linewidth]{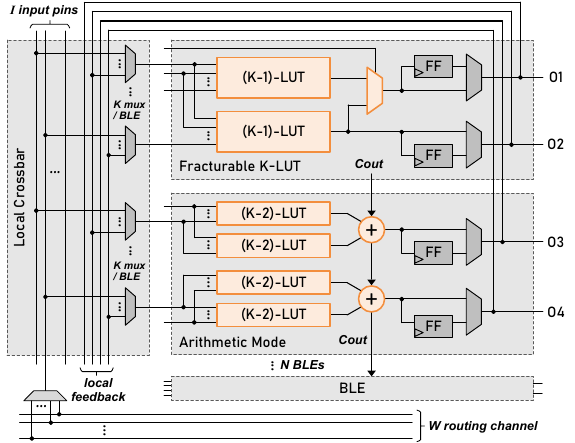}
    \caption{Logic block diagram of the baseline FPGA for VTR architectural exploration.}
    \label{fig:baseline_arch}
\end{figure}

\subsection{Benchmark Workflow}
Unlike many previous FPGA benchmarks \cite{arora:fpl21:koios, Murray2013TitanEL} that provide a fixed Verilog design for every kernel, Kratos provides Python scripts to automatically generate the top-level SystemVerilog module given user-provided design parameters specified in a Python dictionary. These modules contain the pre-implemented kernels we mentioned above and the embedded weights for synthesize.
Thanks to the enhanced synthesis front-end of VTR, our SystemVerilog is human-readable, parameterized, performance-optimized, and extensible. 
After generating the hardware description, the Python scripts generate the necessary flow scripts to run either Intel Quartus Prime or VTR and report the performance and area results. 

All kernels in the Kratos benchmark are parameterized by the dimensions of inputs and weights, as well as sparsity and precision. Sparsity specifies the percentage of zero elements in the weight tensor, and precision specifies the data width of inputs and weights. To simulate unstructured sparsity, we generate the weight matrix with the desired amount of non-zero elements and randomly shuffle their location. 
For precision, Kratos allows any integer data type, but it can be easily extended to support other data formats by changing the hardware description of the MAC unit.  
In addition, all kernels are functionally verified by simulating kernels running on random weights and inputs with Modelsim and comparing results with ground truth
To facilitate large-scale design space exploration, Kratos tools set also provides a batch job script that allows users to define multiple sizes, precisions, and sparsities, and it will launch flows for different combinations.


\section{Evaluation Methodology}

\subsection{Experimental Setup} \label{sec:experiment_setup}
To quantify the efficiency of existing FPGA architectures for unrolled DNNs, we use the Intel Quartus Prime Software Version 22.3 and Arria 10 GX 1150 when running all Kratos benchmarks. 
To conduct FPGA architectural exploration, we use a customized version of VTR\footnote{Our fork from the VTR main branch includes better SystemVerilog support, an option to manually specify the top module, and several bug fixes.} 
As a sanity check to verify the successful parsing of our benchmark through VTR, we compare the resource utilization reported by VTR and Quartus for all kernels and observe $\pm10\%$ variation on average.

The baseline FPGA for our VTR experiments has a Stratix-IV-like architecture using 40 nm technology and is available in the official VTR release. The logic block (LB) diagram of this architecture is shown in Fig. \ref{fig:baseline_arch}, which contains $I = 52$ input pins and a default of $N = 10$ basic logic elements (BLEs). Each BLE contains a LUT with size $K = 6$ and the two outputs can be optionally registered. The BLE can also operate in the fracturable LUT mode where each 6-LUT can be fractured into two 5-LUTs, or the arithmetic mode where the two hard adders receive inputs from four 4-LUTs. To facilitate architectural exploration, we develop a Python-based architecture file generator to automatically modify different LB parameters. 
The area and delay of the modified LB are extracted from COFFE 2.0 \cite{COFFE2} and scaled to 40 nm technology. 
During VTR routing, we set the default router option to perform a binary search to find the minimum routing channel width $W$ required to route the circuit. 

\begin{table}
    \footnotesize	
    \centering
    \caption{Kratos Design Space for Evaluation.}
    \renewcommand{\arraystretch}{1.1}

    \begin{threeparttable}
    \begin{tabular}{ m{1.0cm} m{0.9cm} m{0.8cm}<{\centering} m{1.7cm} m{1.4cm} m{1.7cm}  }
    
        \toprule
          \textbf{Kernel} & \textbf{Unroll \newline Factor\tnote{1}} & \textbf{Size\tnote{1}} & \textbf{Input \newline Dim\tnote{2}} & \textbf{Weight \newline Dim\tnote{3}} & \textbf{Output \newline Dim\tnote{4}} \\
        \bottomrule
        
          gemmt & \multicolumn{1}{c}{RP} & S & $32 \times 32$ & $32 \times 32$ & $32 \times 32$ \\
          \rowcolor{Gray} 
          gemmt & \multicolumn{1}{c}{RP} & L & $128 \times 128$ & $128 \times 128$ & $128 \times 128$ \\

          gemmt & \multicolumn{1}{c}{FU} & S & $16 \times 16$ & $16 \times 16$ & $16 \times 16$  \\
          \rowcolor{Gray} 
          gemmt & \multicolumn{1}{c}{FU} & L & $32 \times 32$ & $32 \times 32$ & $32 \times 32$ \\

          gemms & \multicolumn{1}{c}{RP} & S & $16 \times 16$ & $16 \times 16$ & $16 \times 16$  \\
          \rowcolor{Gray} 
          gemms & \multicolumn{1}{c}{RP} & L & $128 \times 128$ & $128 \times 128$ & $128 \times 128$ \\

          conv1d & \multicolumn{1}{c}{PW} & S & $32 \times 1 \times 64$ & $3 \times 3$ & $30 \times 30 \times 64$ \\
          \rowcolor{Gray} 
          conv1d & \multicolumn{1}{c}{PW} & L & $32 \times 1 \times 64$ & $3 \times 3$ & $30 \times 30 \times 128$ \\

          conv1d & \multicolumn{1}{c}{FU} & S & $32 \times 1 \times 8$ & $3 \times 3$ & $30 \times 30 \times 8$ \\
          \rowcolor{Gray} 
          conv1d  & \multicolumn{1}{c}{FU} & L & $32 \times 1 \times 16$ & $3 \times 3$ & $30 \times 30 \times 16$ \\

          conv2d & \multicolumn{1}{c}{PW} & S & $25 \times 25 \times 32$ & $3 \times 3$ & $23 \times 23 \times 64$ \\
          \rowcolor{Gray} 
          conv2d & \multicolumn{1}{c}{PW} & L & $25 \times 25 \times 64$ & $3 \times 3$ & $23 \times 23 \times 64$ \\

          conv2d & \multicolumn{1}{c}{RP} & S & $8  \times 8  \times 8 $ & $3 \times 3$ & $6 \times 6 \times 8$ \\
          \rowcolor{Gray} 
          conv2d & \multicolumn{1}{c}{RP} & L & $8  \times 8  \times 16$ & $3 \times 3$ & $6 \times 6 \times 16$ \\

          conv2d & \multicolumn{1}{c}{FU} & S & $8  \times 8  \times 4 $ & $3 \times 3$ & $6 \times 6 \times 4$ \\
          \rowcolor{Gray} 
          conv2d & \multicolumn{1}{c}{FU} & L & $8  \times 8  \times 8 $ & $3 \times 3$ & $6 \times 6 \times 8$ \\
          
        \bottomrule
        
    \end{tabular}
    
    \begin{tablenotes}
      \item[1] PW: pixelwise. RP: row-parallel. FU: fully-unrolled. S: small. L: large. 
      \item[2] Format: $m \times n$ for GEMM, $I_W \times I_h \times I_c$ for convolution.
      \item[3] Format: $n \times p$ for GEMM, $F_W \times F_h$ for convolution.
      \item[4] Format: $p \times k$ for GEMM, $O_W \times O_h \times O_c$ for convolution.
    \end{tablenotes}
    \end{threeparttable}
    \label{tab:benchmark_design_space}
\end{table}

\subsection{Design Space}
The Kratos benchmark enables large design space exploration by allowing users to specify arbitrary kernel sizes, as well as sparsity and precision. For our experiments, we use the set of kernel sizes as shown in Table \ref{tab:benchmark_design_space}, which contains two size variants (small and large) for all kernels. 
The convolution kernels have a stride of 1 without padding. 
For every kernel size, we evaluate $10$ evenly spaced sparsity from $0$ to $0.9$, and 4 data precision (1-bit, 2-bit, 4-bit, 8-bit). 
Note that the kernel sizes are chosen to ensure that they can pass the placement and routing under the lowest sparsity level (i.e., no sparsity) and the highest data precision (8-bit). 


\section{Experimental Results}

In this section, we present area and frequency trends of Kratos benchmark circuits to highlight the effect of sparsity and precision.
In addition, we present a proof-of-concept architectural exploration case study to investigate the LUT size for unrolled DNNs, and the potential area savings compared to current general-purpose FPGAs.

\begin{figure}
    \centering
    \includegraphics[width=1\linewidth]{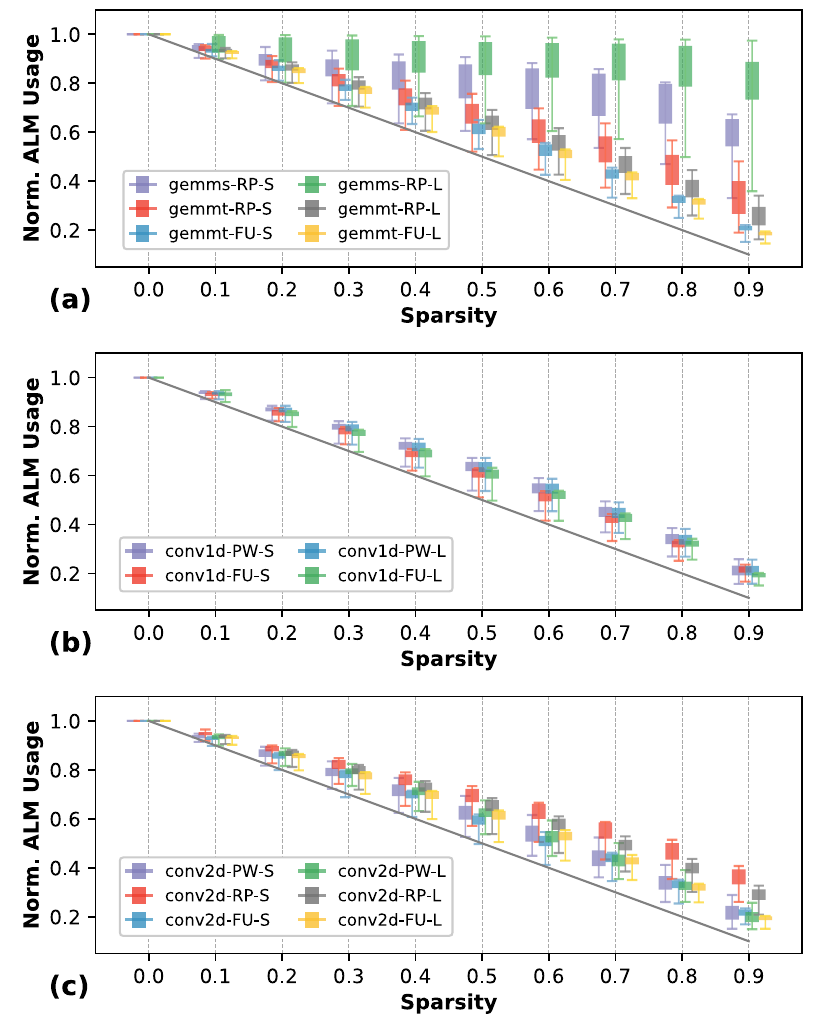}
    \caption{Normalized ALM utilization on Arria 10 vs. sparsity for (a) GEMM, (b) conv1d, and (c) conv2d kernels. The solid black line highlights the ideal trend where the ALM utilization linearly decreases with higher sparsity. }
    \vspace{-5pt}
    \label{fig:alm_quartus}
\end{figure}

\subsection{Area and Frequency Trends on Arria 10}

\textbf{Resource utilization vs. sparsity.} 
Fig. \ref{fig:alm_quartus} shows the normalized adaptive logic module (ALM) utilization vs. sparsity for different Kratos kernels on Arria 10 GX 1150. 
The error bars indicate the range of ALM utilization under different precisions, with the interquartile range marked by filled rectangles. 
Most kernels exhibit a near-ideal linear reduction in ALM utilization with increased sparsity, demonstrating the effectiveness of FPGAs in accelerating unrolled DNNs. 
The row-parallel \textit{gemms} deviates from this trend; at 0.9 sparsity, its ALM utilization is reduced by only 46\% and 31\% for small and large designs, respectively. This is due to \textit{gemms}' structured datapath with delay registers between processing elements, which hampers optimization of zero-weight MAC units. Conversely, the multiply-adder tree implementation prunes zero branches entirely during synthesis, eliminating the need for LUTs and registers as sparsity increases.

\begin{figure}
    \centering
    \includegraphics[width=1\linewidth]{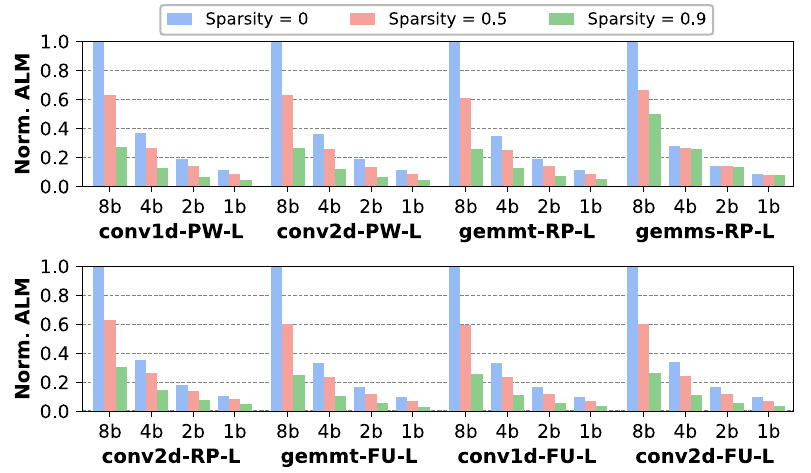}
    \caption{Normalized ALM utilization on Arria 10 vs. precision under different sparsity levels.}
    \vspace{-5pt}
    \label{fig:precision_vs_alm}
\end{figure}

\vspace{.07cm}\noindent\textbf{Resource utilization vs. bit-width.} 
As we decrease bitwidth, area decreases super-linearly as shown in Fig. \ref{fig:precision_vs_alm}.
This is expected because multipliers scale quadratically with bit-width while adders scale linearly and control circuitry remains constant. 
Further inspection of Fig.~\ref{fig:precision_vs_alm} reveals comparative area savings trends from higher sparsity and lower bit-width. 
For instance, when we inspect the 8-bit \textit{conv2d-FU-L} plot, reducing the precision to 4-bits leads to a 2.9-fold decrease in area. Achieving a similar reduction in area for an 8-bit implementation would require high sparsity levels, ranging from 80\% to 90\%.
Recent research is beginning to explore how pruning compares with quantization in terms of accuracy~\cite{kuzmin2023pruning}. 
Together with our hardware efficiency results, this opens the door for more extensive studies on accuracy-efficiency tradeoffs for pruning, quantization, and their combination.

\begin{figure}
    \centering
    \includegraphics[width=1\linewidth, trim=0 0 0 10]{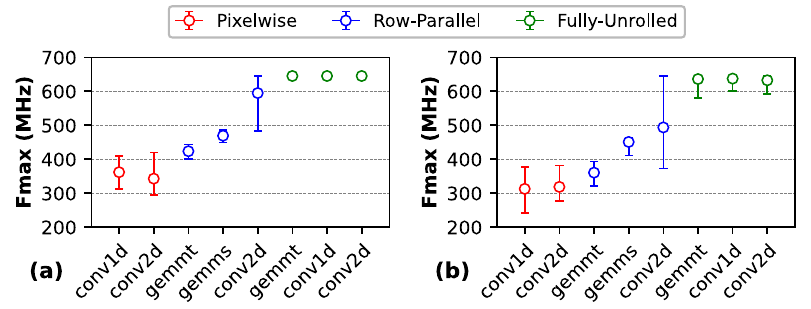}
    \caption{Frequency ranges of (a) small and (b) large Kratos circuits under different sparsity and precision on Arria 10. The circle on each error bar marks the average.}
    \vspace{-5pt}
    \label{fig:freq_quartus}
\end{figure}

\begin{table*}[ht]
    \footnotesize	
    \centering
    \caption{Resource utilization, silicon area, and performance under different LUT sizes, sparsity, and precision.}
    \renewcommand{\arraystretch}{1.1}

    \begin{threeparttable}
    \begin{tabular}{ m{1cm}<{\centering} | m{0.5 cm}<{\centering} | 
    m{0.55cm}<{\centering} m{0.7cm}<{\centering} m{0.8cm}<{\centering} | 
    m{0.55cm}<{\centering} m{0.7cm}<{\centering} m{0.8cm}<{\centering} | 
    m{0.55cm}<{\centering} m{0.7cm}<{\centering} m{0.8cm}<{\centering} | 
    m{0.55cm}<{\centering} m{0.7cm}<{\centering} m{0.8cm}<{\centering} | 
    m{0.55cm}<{\centering} m{0.7cm}<{\centering} m{0.8cm}<{\centering} | 
    m{0.55cm}<{\centering} m{0.7cm}<{\centering} m{0.8cm}<{\centering}    }
    
    \toprule
      \multirow{3}{*}{Sparsity} & \multirow{3}{*}{$K$ \tnote{1}} & \multicolumn{9}{c|}{Precision = 8-bit} & \multicolumn{9}{c}{Precision = 4-bit} \\ 
      
      \cline{3-20}
      
      & & \multicolumn{3}{c|}{gemmt-RP-S} & \multicolumn{3}{c|}{conv1d-PW-S} & \multicolumn{3}{c|}{conv2d-PW-S} & \multicolumn{3}{c|}{gemmt-RP-S} & \multicolumn{3}{c|}{conv1d-PW-S} & \multicolumn{3}{c}{conv2d-PW-S} \\
      
      \cline{3-20}
      
      & & $k$LBs & Area ($mm^2$) & Fmax (MHz) & $k$LBs & Area ($mm^2$) & Fmax (MHz) & $k$LBs & Area ($mm^2$) & Fmax (MHz) & $k$LBs & Area ($mm^2$) & Fmax (MHz) & $k$LBs & Area ($mm^2$) & Fmax (MHz) & $k$LBs & Area ($mm^2$) & Fmax (MHz)  \\ 
      
      \midrule

      \multirow{4}{*}{$0\%$} 
& 3 & 1.13 & \tb{1.88} & 93.1 & 13.4 & \tb{22.4} & 44.0 & 20.4 & \tb{34.0} & 38.6 & 0.97 & \tb{1.62} & \tb{177.9} & 11.3 & \tb{18.9} & 113.6 & 17.1 & \tb{28.4} & 87.7 \\ 
& 4 & 1.06 & 2.18 & \tb{124.7} & 12.8 & 26.2 & 45.5 & 19.1 & 39.3 & \tb{41.9} & 0.97 & 2.0 & 170.4 & 11.3 & 23.3 & 116.4 & 17.0 & 35.0 & 79.1 \\ 
& 5 & 1.06 & 2.67 & 102.0 & 12.8 & 32.2 & \tb{53.5} & 19.2 & 48.3 & 40.0 & 0.97 & 2.45 & 172.0 & 11.3 & 28.5 & \tb{118.9} & 17.0 & 42.9 & 81.4 \\ 
& 6 & 1.06 & 3.62 & 114.5 & 12.8 & 43.7 & 46.5 & 19.2 & 65.6 & 37.7 & 0.97 & 3.32 & 167.8 & 11.4 & 38.9 & 118.9 & 17.1 & 58.5 & \tb{93.4} \\ 

      \midrule

      \multirow{4}{*}{$50\%$} 
& 3 & 0.66 & \tb{1.11} & \tb{146.9} & 8.05 & \tb{13.4} & 72.6 & 12.4 & \tb{20.6} & \tb{49.6} & 0.5 & \tb{0.83} & \tb{179.8} & 6.01 & \tb{10.0} & 118.7 & 9.07 & \tb{15.1} & \tb{92.9} \\ 
& 4 & 0.63 & 1.29 & 143.8 & 7.58 & 15.6 & \tb{73.7} & 11.5 & 23.6 & 46.5 & 0.5 & 1.02 & 179.8 & 6.01 & 12.3 & 106.0 & 9.06 & 18.6 & 87.2 \\ 
& 5 & 0.63 & 1.59 & 144.9 & 7.57 & 19.1 & 65.4 & 11.4 & 28.8 & 48.2 & 0.5 & 1.25 & 172.3 & 6.01 & 15.1 & 106.8 & 9.06 & 22.8 & 88.6 \\ 
& 6 & 0.63 & 2.15 & 135.6 & 7.57 & 25.9 & 70.0 & 11.4 & 39.0 & 48.6 & 0.5 & 1.7 & 173.4 & 6.01 & 20.6 & \tb{119.5} & 9.06 & 31.0 & 73.8 \\ 

      \midrule

      \multirow{4}{*}{$90\%$} 
& 3 & 0.21 & \tb{0.35} & \tb{169.4} & 2.74 & \tb{4.56} & 100.1 & 4.5 & \tb{7.48} & 76.9 & 0.1 & \tb{0.17} & 179.9 & 1.25 & \tb{2.09} & 124.0 & 2.02 & \tb{3.37} & 106.0 \\ 
& 4 & 0.21 & 0.43 & 165.8 & 2.58 & 5.31 & 109.1 & 4.13 & 8.48 & \tb{99.0} & 0.1 & 0.21 & 179.1 & 1.24 & 2.55 & \tb{134.4} & 1.93 & 3.96 & 105.3 \\ 
& 5 & 0.21 & 0.52 & 166.5 & 2.58 & 6.51 & \tb{125.3} & 4.08 & 10.3 & 86.1 & 0.1 & 0.26 & \tb{183.5} & 1.24 & 3.12 & 118.0 & 1.92 & 4.83 & 103.7 \\ 
& 6 & 0.21 & 0.7 & 165.9 & 2.58 & 8.83 & 101.1 & 4.05 & 13.8 & 92.4 & 0.1 & 0.35 & 181.6 & 1.24 & 4.24 & 132.4 & 1.92 & 6.56 & \tb{108.7} \\ 
 
      \bottomrule
        
    \end{tabular}
    
    \begin{tablenotes}
        \item[1] For $K = {3,4,5,6}$, the maximum channel widths required to route all designs are $W = {102, 96, 90, 90}$, which gives a tiles area of ${1664 um^2, 2053 um^2, 2520 um^2, 3420 um^2}$ from COFFE~\cite{COFFE2} after normalizing to 40 nm technology.
    \end{tablenotes}
    \end{threeparttable}
    \label{tab:lutsize_vs_area}
\end{table*}

\vspace{.0cm}\noindent\textbf{Critical Path Delay.} 
Fig. \ref{fig:freq_quartus} shows the frequency ranges of small and large Kratos kernels on Arria 10.
There is a clear trend in favor of higher unrolling factors, with our heavily-pipelined fully-unrolled designs reaching the maximum frequency supported on Arria 10.
In this case, the unrestricted fmax can reach 1GHz by Quartus timeing report, and the restricted fmax can reach over  600Mhz.
Conversely, the row-parallel and pixelwise implementations suffer from higher critical path delays within the control and buffering circuitry but are still capable of reaching frequencies 300--600~MHz.
The high speeds attainable with unrolled DNNs, combined with their direct area savings from fine-grained sparsity and reduced bit-width motivate further investigation of new FPGA architectures to enable larger DNN deployments.

\subsection{Architectural Exploration Case Study}
%
Using the baseline architecture described in Section \ref{sec:experiment_setup}, we conduct a case study to find the optimal LUT size for unrolled DNNs. 
We evaluate four LB architectures whose LUT sizes $K$ vary from $3$ to $6$. 
For each different LUT size, we determine the corresponding number of LB input pins $I$ with $N = 10$ basic logic elements based on the empirical equation $I=\frac{K}{2}(N+1)$ from prior work~\cite{Ahmed2000TheEO}. 
We use VTR to evaluate the four architectures on three kernels \textit{gemmt-RP-S}, \textit{conv1d-PW-S}, \textit{conv2d-PW-S} from Table \ref{tab:benchmark_design_space} as these designs balance silicon footprint and data throughput. While fully unrolled designs offer the best clock frequency, practical constraints and the need to run multiple experiments necessitated choosing smaller, more manageable designs in this initial exploration.

For every architecture, we extract the maximum routing channel width $W$ reported by VTR that can fit all designs, which is then passed to COFFE~\cite{COFFE2} to compute the LB area (including routing) with a given $(K, I, W)$. 
The total silicon area of a kernel is then calculated by multiplying the LB utilization and the LB area. All reported results are averaged over 3 runs using different random seeds 

\begin{figure}
    \centering
    \includegraphics[width=1\linewidth]{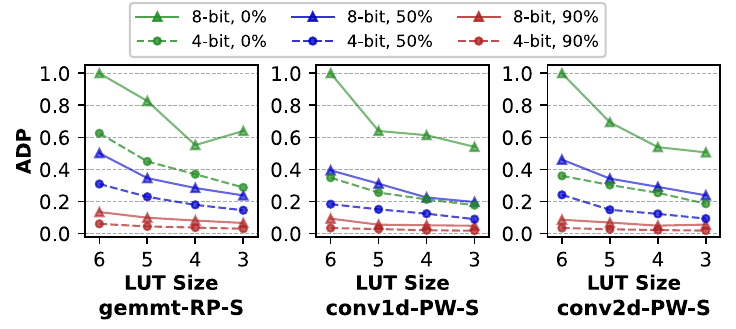}
    \caption{Normalized area-delay product (ADP) for the Kratos circuits in Table \ref{tab:lutsize_vs_area}.}
    \vspace{-5pt}
    \label{fig:lutsize_ADP}
\end{figure}

The experiment results summarized in Table \ref{tab:lutsize_vs_area} show potential savings of $\sim$2$\times$ when reducing the LUT size from 6 (default in most Intel FPGAs) to 3.
This comes with a 10--20\% degradation in critical path delay for 8-bit kernels, whereas a small improvement is observed for most 4-bit designs.
We hypothesize that smaller 4-bit MAC units are more likely to fit within a single logic block, even with $K = 3$, compared to 8-bit counterparts.
When optimizing the FPGA device for area-delay product, Fig.~\ref{fig:lutsize_ADP} favors the smallest LUT size ($K=3$) except for one circuit (\textit{gemmt-RP-S} with 8-bit precision and no sparsity). 
This strongly indicates the superiority of smaller LUTs for unrolled DNN implementations. 

\section{Conclusions and Future Work}
Motivated by the efficiency advantages of unrolled DNNs, we created a benchmark suite to enable the architectural exploration of new programmable hardware devices for accelerating unrolled DNNs.
Our empirical analysis shows that unrolled DNNs on FPGA can run at very high speed, can can significantly benefit from improvements in efficiency with fine-grained sparsity and reduced arithmetic precision --- two properties that are not easily attainable with conventional DNN accelerators.
Furthermore, we performed an architectural case study to reveal $\sim$2$\times$ possible area savings from exploring the optimal LUT size of contemporary FPGA architectures to suit unrolled DNNs better.

While we can't optimize a general-purpose FPGA solely for unrolled DNNs, future work can explore integrating specialized bit-programmable fabrics within general-purpose FPGAs or creating new bit-programmable devices specifically for unrolled DNNs. One goal of Kratos is to inspire research on new programmable architectures that are much more efficient (e.g., $10-100\times$) than current FPGAs, maintaining linear and quadratic efficiency scaling with sparsity and low precision, respectively.
Although the size of unrolled DNNs that can fit on current FPGAs is small, future works can explore algorithmic optimizations such as weight sharing and time-domain multiplexing to drastically increase the capacity of unrolled DNNs that can fit on the target bit-programmable device. For example, with weight sharing \cite{Lan2019ALBERTAL}, there can be one large unrolled layer that is shared throughout the DNN, and a small accelerator for “adapter” layers that run much slower. Although time-domain multiplexing (investigated by Tabula Inc.) has been proven challenging for general-purpose FPGAs, it could be a really good fit for programmable devices targeting unrolled DNNs with a much simpler CAD flow due to domain specialization, and can achieve multiplicative efficiency on mapping larger unrolled DNNs. We believe that the Kratos benchmark is the first and valuable step to begin investigating unrolled DNNs on programmable architectures, which is a promising research direction because it addresses an open problem in DNN research: how to fully leverage fine-grained unstructured sparsity and mixed precision effectively---something that current accelerators cannot handle. 

\clearpage 
\bibliographystyle{IEEEtran}
\bibliography{references}

\end{document}